\documentclass[%
 reprint,
 amsmath,amssymb,
 aps,
floatfix
]{revtex4-2}

\usepackage{graphicx}
\usepackage{dcolumn}
\usepackage{bm}

\begin{document}

\preprint{APS/123-QED}

\title{Femtosecond laser written low-loss multiscan waveguides in fused silica}

\author{N.N. Skryabin}
\email{nikolay.skryabin@phystech.edu}
\affiliation{%
 Quantum Technology Centre and Faculty of Physics, M.V. Lomonosov Moscow State University, 1 Leninskie Gory Street, Moscow 119991, Russia}%
  
\author{S.A. Zhuravitskii}
\affiliation{%
 Quantum Technology Centre and Faculty of Physics, M.V. Lomonosov Moscow State University, 1 Leninskie Gory Street, Moscow 119991, Russia}%
 
\author{I.V. Dyakonov}
\affiliation{%
 Quantum Technology Centre and Faculty of Physics, M.V. Lomonosov Moscow State University, 1 Leninskie Gory Street, Moscow 119991, Russia}%
\affiliation{%
 Russian Quantum Center, 30 Bolshoy bul'var building 1, Moscow 121205, Russia}%

\author{A.A. Kalinkin}
\affiliation{%
 Quantum Technology Centre and Faculty of Physics, M.V. Lomonosov Moscow State University, 1 Leninskie Gory Street, Moscow 119991, Russia}%
\affiliation{Laboratory of Quantum Engineering of Light, South Ural State University (SUSU), 76 Prospekt Lenina, Chelyabinsk 454080, Russia}%

\author{S.S. Straupe}
\affiliation{%
 Quantum Technology Centre and Faculty of Physics, M.V. Lomonosov Moscow State University, 1 Leninskie Gory Street, Moscow 119991, Russia}%
\affiliation{%
 Russian Quantum Center, 30 Bolshoy bul'var building 1, Moscow 121205, Russia}%
 
\author{S.P. Kulik}
\affiliation{%
 Quantum Technology Centre and Faculty of Physics, M.V. Lomonosov Moscow State University, 1 Leninskie Gory Street, Moscow 119991, Russia}%
\affiliation{Laboratory of Quantum Engineering of Light, South Ural State University (SUSU), 76 Prospekt Lenina, Chelyabinsk 454080, Russia}%

\date{\today}

\begin{abstract}
We report low-loss multiscan waveguides fabricated in fused silica using femtosecond laser writing technology. The multiscan principle allows to tailor the writing regime to excel at key features of any integrated photonic platform: coupling losses and propagation losses. We optimized the writing parameters for different sizes of square-shaped waveguides and reached the mode overlap value with a standard single-mode optical fiber of above 98.8\% and demonstrated very low coupling losses of 0.2 dB/facet on average. Propagation losses in the fabricated waveguides amounted to 0.07 dB/cm. We applied the developed recipe to the fabrication of a fiber-coupled 25-channel interferometer with total insertion losses below 1 dB. The findings of this work are of interest for broad range of applications and in particular for optical information processing and quantum photonics.

\end{abstract}

\maketitle


\section{\label{sec:Intro}Introduction}

Femtosecond laser writing (FLW) technology is the unrivaled tool for rapid prototyping of integrated photonic devices \cite{ Tan2016review, Tan2021review, Li2023review, GrossWithford2015review} in various dielectric materials due to its low cost and simplicity. FLW has found a wide range of applications in astrophotonics \cite{GrossWithford2015review}, optical communications \cite{GrossWithford2015review, Cai2022review}, topological photonics \cite{Szameit2013, Maczewsky2017, Arkhipova2023, Ren2023}, optofluidics \cite {Sugioka2014review, Choi2020review}, and quantum photonics \cite{Meany2015review, CorrielliCrespiOsellame2021}, where a key element is a multichannel interferometer. Moreover, it was this technology that was at the origin of the use of integrated interferometers in quantum optical experiments \cite{Marshall2009, Sansoni2010}, which was followed by a series of works aimed at the development of single-qubit \cite{Crespi2011, Corrielli2014, Heilmann2014} and multi-qubit gates \cite{Meany2016, Zeuner2018, Zhang2019, Li2020, Li2022, Skryabin2023}, and also at carrying out quantum walks \cite{Owens2011, Sansoni2012, Crespi2013_2, Giuseppe2013, Poulios2014, Tang2018, Ehrhardt2021} and boson sampling experiments \cite{Crespi2013, Tillmann2013, Spagnolo2013, Spagnolo2014, Bentivegna2015, Anton19, Hoch2022}. Nevertheless, in all of these works, only a few photons (2 -- 4) were used, thus, the moderate insertion loss (IL) $\sim$ 3 -- 10 dB of the chips was not a significant concern. The propagation losses (PLs) were $\sim$ 0.2 -- 1.0 dB/cm, while the coupling losses (CLs) were $\sim$ 0.7 -- 3 dB/facet at 800 -- 900 nm. 

Since transmission through the chip has an exponential effect on multiphoton events, IL becomes a decisive parameter of an integrated platform, determining the scaling to more complex experiments with a larger number of photons.  Consequently, large-scale experiments tend to use low-loss bulk optics and fiber optics \cite{Zhang2016, Wang2017, Wang2019, Li2020Graph, Madsen2022, cao2023photonic} or state-of-the-art silicon photonics \cite{Vigliar2021, Maring2024, Chen2024}, where the total IL is $\sim$ 3 -- 3.5~dB. Recently, borosilicate glass-based FLW chips with lower IL of $\sim$ 2 -- 3~dB have been developed \cite{Albiero2022, Pont2022, Pont2022_2, Albiero2024}, where the PL of 0.13 dB/cm \cite{Albiero2022, Albiero2024} and the CL of 0.2~dB/facet \cite{Albiero2024} were achieved in the NIR range of wavelengths. 

One way to further enhance the FLW platform is to optimize waveguides in fused silica. These waveguides, depending on the pulse energy, pulse duration and focusing conditions, are divided into two types according to the modification characteristics of the material \cite{Itoh2006, Poumellec11, Mishchik2013}. Type-I is associated with local melting and has an isotropic refractive index change  $\Delta n \sim 10^{-3}$ and exhibit 0.1~dB/cm PL \cite{Maczewsky2017, Arkhipova2023, Ren2023} at around 800~nm. Such waveguides possess fundamental eigenmodes with mode field diameter (MFD) of the order of 10-20~$\mu$m due to the low refractive index contrast, which leads to mode mismatch with the conventional optical fibers. On the contrary type-II waveguide fabrication is accompanied by the formation of nanogratings and has an anisotropic refractive index change $\Delta n \sim 10^{-3} - 10^{-2}$ and moderate propagation loss $0.3 - 0.8$ dB/cm \cite{Eaton2011, Fernandes2011, Tillmann2013, Dyakonov2018}, which has been optimized to 0.15 dB/cm at the wavelength of 1550 nm \cite{Amorim2017, Tan21}. Recently, a record low PL of 0.07 dB/cm has been demonstrated at 1550 nm with a total chip IL of 1.07 dB \cite{Tan22}. For a wavelength of 900~nm, the primary challenge is also to minify large CL $\sim$ 1~dB/facet \cite{Zeuner2018, Skryabin2023}. 

Here, we report the design and test of square shape waveguides with improved properties in fused silica. The waveguides were written using multiscan-with-a-shift method, which consists of writing a series of consecutive mutually shifted type-I waveguides with minor transversal step between them. Similar techniques have recently been demonstrated to reduce PL \cite{Tan22}, bending loss (BL) \cite{Tan22, Lee2021, Ross-Adams2024}, CL \cite{Heilmann18, Fernandez2022, Ehrhardt23, Ross-Adams2024} and waveguide mode manipulation \cite{Sun2022, Wang2024}. However, in CL optimization insufficient attention has been paid to the shape, size or writing depth of the resulting waveguides. We performed theoretical simulations of the mode overlap with an optical fiber based on the type-I waveguide sizes that are feasible to form at different writing depths using appropriate pulse energy and deduced the optimal writing parameters. We then fabricated and characterized square waveguides with the size optimized to match the input fiber and measured the CL of 0.2~dB/facet on average. The waveguides also exhibited relatively low PL equal to 0.07~dB/cm. Next, we tested the homogeneity of the waveguides by comparing the repeatability of directional couplers (DCs) and Mach-Zehnder interferometers (MZIs) for two multiscan waveguide writing orders. Finally, we fabricated a 25-channel interferometer with a total IL below 1 dB, which is suitable for multiphoton boson sampling experiments.

\section{\label{sec:Methods}Methods}

The waveguides were fabricated by FLW technique in fused silica samples (JGS1 glass material) with different lengths of 3 -- 10~cm. Power stabilized femtosecond pulses from a frequency doubled ytterbium fiber laser (Avesta Antaus) with the central wavelength of 515~nm, pulse duration of 270~fs, pulse energies in the range of 14 -- 70~nJ and the repetition rate of 1~MHz were focused with an aspheric lens (NA = 0.55) 100 -- 1000 $\mu$m below the surface through a 450~$\mu$m cover glass used to compensate for spherical aberrations. A variable beam expander was used to increase the laser beam diameter in order to completely fill the entrance aperture of the focusing lens. The sample was translated relative to the focal spot with a high-precision stage (AeroTech ANT) at a writing speed of 1 -- 6~mm/s. Circular polarization of the laser beam was set by a quarter-wave plate $\lambda/4$. The FLW setup scheme is provided in Appendix~\ref{app:setups_schemes}.

The input and output facets of the sample were polished to optical quality after the writing process was completed. Top and facet microscope images of the written waveguides were taken using an optical microscope (Carl Zeiss Axio Scope A1) with a 40X optical objective in the bright field (BF) or in the differential interference contrast (DIC) regimes. The optical properties of the waveguides were characterised at 920~nm using a diode laser (Toptica CTL950). The laser radiation was butt-coupled into the sample through a single cleaved fiber pigtail (Nufern PM780-HP). The slow axis of the fiber was aligned with the horizontal linear polarization (H-polarization). The sample was placed on a 6-axis mechanical alignment stage (Luminos) during the characterization. The output radiation was focused with an aspheric lens (F = 7.5 mm) onto the CMOS camera (Beamage-4M) to capture the near-field spatial profile of the waveguide's eigenmode or collimated onto the power meter to measure transmittance. The fiber array was used to collect the output light during the characterization of the 25-channel interferometer. The characterization setup scheme, detailed IL decomposition and MZI's internal phase evaluation are provided in Appendix~\ref{app:setups_schemes}.

\section{\label{sec:Experiment}Experiment and Results}

\subsection{\label{sec:E1}Type-I single scan waveguides}

\begin{figure*}[ht]
\centering
\includegraphics[width=1.0\linewidth]{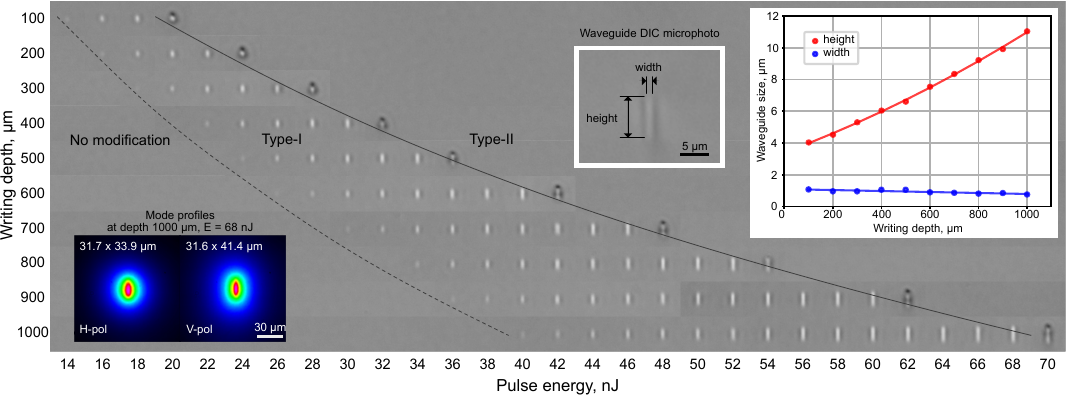}
\caption{The stitched facet view micrographs of single scan waveguides written in the depth range from 100 to 1000~$\mu$m with different energies, writing speed is 1~mm/s. The picture illustrates the type-I waveguides energy range from modification threshold (dash line) to the transition to type-II waveguides (solid line) at each depth. The sizes of the waveguides in the pictures do not correspond to the real ones, since the pictures were taken in BF regime with a closed aperture for better visibility of the waveguides. The inset on the top right shows a waveguide photographed in a DIC regime, and a plot of the actual sizes of Type-I boundary waveguides (at maximum energy preceding the transition to the Type-II waveguides) for different depths. The lower left inset shows the mode profiles of the waveguide written at 1000~$\mu$m depth and with 68~nJ pulse energy.}
\label{fig:fig1}
\end{figure*}

The first step was to write a series of single scan waveguides in the depth range of 100 to 1000 $\mu$m with different energies starting below the modification threshold and spanning up to the transition to the type-II waveguide formation. The writing speed was fixed at 1~mm/s. Facet view micrographs of the waveguides were obtained in the BF regime and are presented in the Fig.~\ref{fig:fig1}. The quadratic change of the modification threshold and type-II modification energy depending on the depth is due to spherical aberrations \cite{Bukharin2017}. As a result, an expansion of the energy range for type-I waveguides with depth is observed. For each depth, a type-I waveguide written at maximal energy was selected because it features a maximal refractive index contrast. The transversal sizes of the selected waveguides were measured using micrographs produced in the DIC regime. The dependence of the sizes on depth is shown in Fig.~\ref{fig:fig1} in the inset on the upper right. The height of the waveguide also increases quadratically with the writing depth, but the width remains almost constant and is approximately equal to $1~\mu$m. The refractive index contrast of these waveguides is quite low, for example, as evidenced in Fig.~\ref{fig:fig1} in the inset in the bottom left corner the MFDs of the order of 30 -- 40~$\mu$m range for the waveguide written at the depth of 1000~$\mu$m and with the energy of 68~nJ. Whereas most of these waveguides do not support any guiding modes.

\subsection{\label{sec:E2}Simulations of the fiber coupling efficiency with rectangular waveguides}

\begin{figure}[h]
\centering
\includegraphics[width=1.0\linewidth]{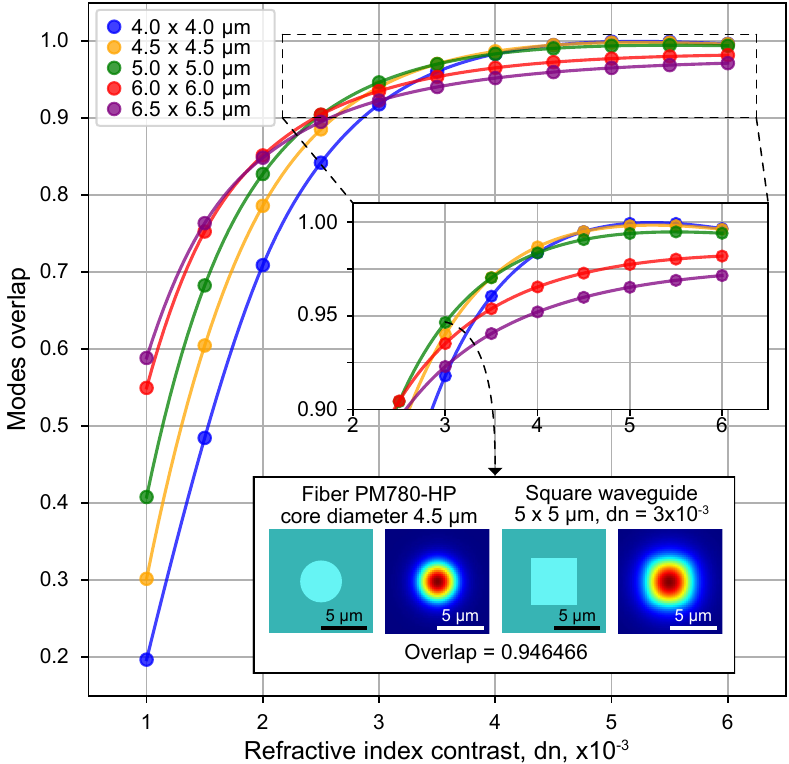}
\caption{Mode overlap of optical fiber with square waveguides of different sizes from 4~$\mu$m to 6.5~$\mu$m, depending on the refractive index contrast $1 - 6 \times 10^{-3}$. The inset below shows examples of waveguide and fiber cores, their modes and the calculated overlap.}
\label{fig:fig2}
\end{figure}

The next step was to calculate the mode matching of rectangular waveguides (assuming that such waveguides could be obtained using the multiscan method) with heights and widths from 2~$\mu$m to 8~$\mu$m and a standard Nufern PM780-HP optical fiber using the Overlap function of the Ansys Lumerical MODE software. Since the resulting refractive index contrast $\Delta n$ of the waveguides can not be precisely controlled, $\Delta n$ in range from $1 \times 10^{-3}$ to $6 \times 10^{-3}$ were considered. As expected, the best match is achieved with square-shaped waveguides. The results for experimentally available sizes are shown in the Fig.~\ref{fig:fig2}. It indicates that for a given $\Delta n$ there exists a waveguide size at which the overlap is maximum. As the $\Delta n$ increases, the overlap maximum shifts towards the smaller waveguides. The best overlap higher than 99$\%$ is achieved if high $\Delta n$ values of the order of $5 \times 10^{-3}$ are set. However even for $\Delta n$ of $3 \times 10^{-3}$ it is possible to achieve an overlap of about 95$\%$ for a waveguide with sizes of $5 \times 5$~$\mu$m. Detailed description of simulation and results are provided in Appendix~\ref{app:simulation}.

\subsection{\label{sec:E3}Square shape multiscan waveguides}

\begin{figure}[h!]
\centering
\includegraphics[width=1.0\linewidth]{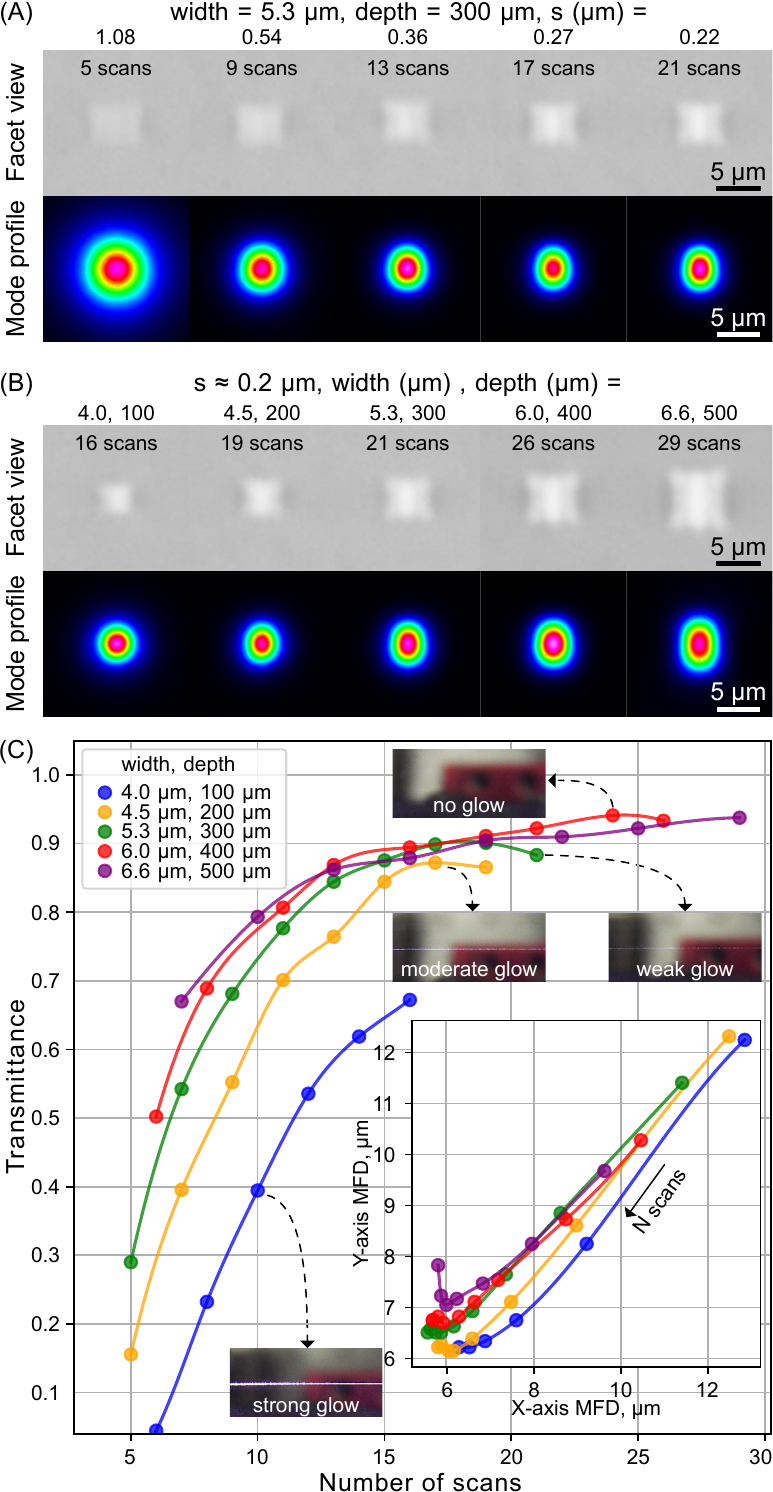}
\caption{(a) Facet view micrographs (in DIC regime) of multiscan waveguides with fixed width of 5.3~$\mu$m and different shifts at a depth of 300~$\mu$m, accompanied by corresponding mode profiles. Number of scans $N$ is indicated at the top of facet view micrographs. (b) Same pictures for multiscan waveguides with a shift s $\approx$ 0.2~$\mu$m at different depths of 100 -- 500~$\mu$m with corresponding widths. (c) Dependence of the waveguides transmittance at different writing depths on the number of scans. The inset graph shows the effect of the number of scans on MFDs for each depth. The arrow shows the direction of increase in the number of scans. The inset pictures show the disappearance of the waveguide glow with increasing depth.}
\label{fig:fig3}
\end{figure}

Next, using the multiscan-with-a-shift method, multiscan waveguides were written at depths of 100, 200, 300, 400 and 500~$\mu$m in a 3~cm long sample applying different number of scans $N$. The waveguides width $w_m$ were chosen to be equal to the height of single scan Type-I boundary waveguides in order for the waveguide shape to be close to square. To kept the waveguide width $w_m$ constant while varying number of scans $N$ the shift $s$ between the scans was set according to $w_m=w_s + (N-1) \times s$, where $w_s$ = 1~$\mu$m is the width of the single scan waveguide. Since the widths of the waveguides are different with depth, the range of the number of scans was varied to keep the shift in the range 0.2~$\mu$m $ \leqslant s \leqslant w_s$ = 1~$\mu$m. Writing energies varied slightly around energies of single scan Type-I boundary waveguides. All writing parameters are listed in Table~\ref{tab:table1}. The ordering of the consecutive writing scans was determined according the principle "from the center to the sides". Detailed comparison of scans ordering principles are provided in Appendix~\ref{app:writing_order}. 

Facet view micrographs of waveguides with a width of 5.3~$\mu$m at a depth of 300~$\mu$m with increasing number of scans $N = 5 - 21$ and corresponding mode profiles are shown in Fig.~\ref{fig:fig3}a. Same pictures for waveguides with a shift $s$ $\approx$ 0.2~$\mu$m at different depths of 100 -- 500~$\mu$m with corresponding widths are shown in Fig.~\ref{fig:fig3}b. The inhomogeneous internal structure of the waveguides is evident, which also affects the symmetry of the mode profiles. We noted that occasionally for a large number of scans $N > 20$ spontaneous microcrack formation was observed during the end-face polishing process, as was previously described in \cite{Lee2021}.

\begin{table}[h!]
\caption{\label{tab:table1}Multiscan waveguides writing parameters.}
\begin{ruledtabular}
\begin{tabular}{cccc}
depth, $(\mu m)$ & $w_m$, ($\mu m$) & $N$ & $E$, $nJ$ \\
\hline
100 & 4.0 & 4 - 16 & 15 - 17 \\
200 & 4.5 & 5 - 19 & 19 - 21 \\
300 & 5.3 & 5 - 21 & 23 - 25 \\
400 & 6.0 & 6 - 26 & 27 - 29 \\
500 & 6.6 & 7 - 29 & 32 - 34 \\
\end{tabular}
\end{ruledtabular}
\end{table}

The laser radiation was injected into the waveguides via the butt-coupled fiber to test the transmission properties. Transmission through each waveguide was measured and the mode profiles were imaged. The dependencies of the transmittance for waveguides at different writing depths on the number of scans are shown in Fig.~\ref{fig:fig3}c. The inset graph shows the effect of the number of scans on MFDs. It can be seen that as the number of scans increases, the transmittance through the waveguides increases significantly and reaches saturation. A similar trend is observed as the MFDs shrink as a result of the increasing refractive index contrast, which leads to the improved mode matching. The high transmittance values above 90\% (IL = 0.5 dB) were registered in the waveguides written at the depths of 400~$\mu$m and 500~$\mu$m with $N > 20$~scans, despite the fact that their modes are slightly larger than that of waveguides at the depth of 200~$\mu$m, where the minimum achieved MFDs is $5.8 \times 6.2$~$\mu$m (while the fiber mode MFDs is $5.8 \times 5.8$~$\mu$m). This indicates that the PL is low at the depths of 400~$\mu$m and 500~$\mu$m. Another indirect confirmation of the smaller PL is the fading of the waveguide glow with increasing writing depth (see the inset images in the Fig.~\ref{fig:fig3}a). The waveguide glow is inherent to type-II waveguides, where strong scattering occurs on nanogratings leading to additional losses.

\subsection{\label{sec:E4}Waveguides optimization and characterization}

\begin{table}[h]
\caption{\label{tab:table2}Multiscan waveguides writing parameters for a more precise optimization.}
\begin{ruledtabular}
\begin{tabular}{ccccc}
s, ($\mu m$) & $w_m$, ($\mu m$) & $N$ & depths, $(\mu m)$ & $E$, $nJ$ \\
\hline
0.2; 0.25 & 4.6; 4.5 & 19; 15 & 200 - 500 & 19 - 30 \\
0.2; 0.25 & 5.0; 5.0 & 21; 17 & 300 - 600 & 22 - 34 \\
0.2; 0.25 & 5.4; 5.5 & 23; 19 & 400 - 700 & 27 - 37 \\
0.2; 0.25 & 5.8; 6.0 & 25; 21 & 400 - 700 & 28 - 38 \\
\end{tabular}
\end{ruledtabular}
\end{table}

\begin{figure}[h!]
\centering
\includegraphics[width=1.0\linewidth]{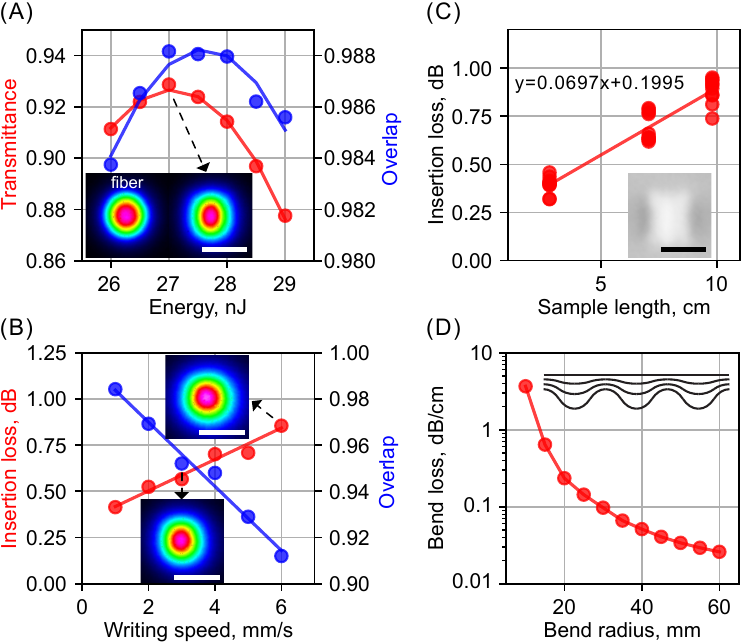}
\caption{(a) Dependence of the waveguides transmittance (red) and mode overlap (blue) on the writing energy. The inset pictures show fiber mode and waveguide mode profiles at energy 27 nJ. (b) Dependence of IL (red) and modes overlap (blue) on the writing speed. The inset picture shows waveguide mode profiles at writing speed 3 and 6 mm/s. (c) Cut–back loss measurement result. The inset picture shows facet-view microphotograph of the waveguide. (d) BL of waveguides for different bending radii. The inset picture shows scheme of the curved waveguides written for BL measurement. The scale bar of 5~$\mu$m is the same for all pictures. All waveguides are written with default optimal parameters (writing depths of 400~$\mu m$, speed $v$ = 1~mm/s, pulse energy $E_p$ = 27~nJ, $w_m = 5.5$~$\mu$m, $s$ = 0.25~$\mu$m, $N$ = 19).} 
\label{fig:fig4}
\end{figure}

The next step was a more precise optimization in the space of fabrication parameters: the waveguide width $w_m$, the shift $s$, the number of scans $N$, the writing depth, and the writing energy $E_p$. Values $s = 0.2$~$\mu$m and $s = 0.25$~$\mu$m were fixed, where the transmittance saturation was observed as described in the previous section. Waveguide widths of 4.5 - 6~$\mu$m were chosen since their corresponding modes are close to the fiber mode. The energy and depth ranges were chosen in order to produce a nearly square waveguide. All parameters are listed in Table~\ref{tab:table2}. We observed the best performance in waveguides at the depth of 400~$\mu$m with $w_m = 5.5$~$\mu$m, $s = 0.25$~$\mu$m and $N$ = 19. The dependence of the waveguides transmittance and mode overlap on writing energy is shown in Fig.~\ref{fig:fig4}a. At the writing energy $E_p$ = 27 nJ the transmittance above 94$\%$ and the mode overlap above 98.8$\%$ were achieved. These parameters deteriorate slightly if the writing speed is increased (see Fig. \ref{fig:fig4}b).

Next, the waveguides losses were measured using the cut-back method using 3 samples of different lengths: 3, 6, 10~cm. Fig.~\ref{fig:fig4}c presents the measurement results. The PL value of 0.07~dB/cm and the CL is 0.2~dB on average were estimated from fits. The dependence of BL on the bending radius (see the Fig.~\ref{fig:fig4}d) follows from the transmission measurement carried out with the curved waveguides consisting of six s-bends with different radii of curvature. We measured additional BL values of 0.1~dB/cm and 0.026~dB/cm for the radius of curvature of 30~mm and 60~mm respectively.

\subsection{\label{sec:E5}Homogeneity of waveguides}

\begin{figure}[h!]
\centering
\includegraphics[width=1.0\linewidth]{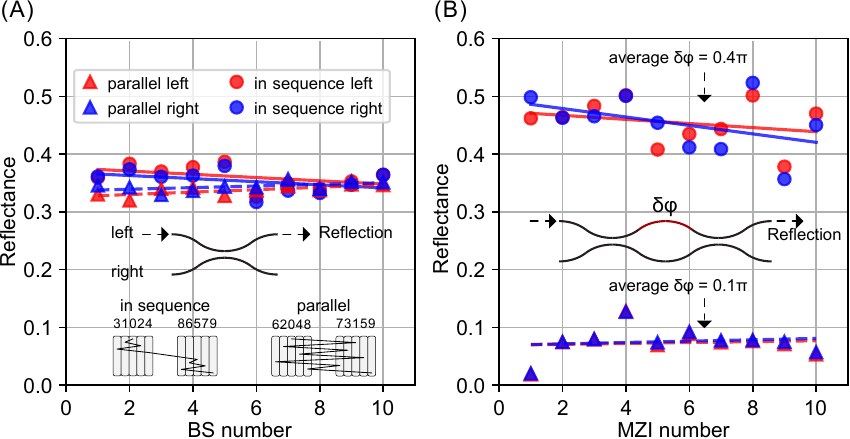}
\caption{(a) Reflectance statistics of DCs where multiscan waveguides are written either sequentially (circle mark) or in parallel (triangle mark). The laser radiation was injected into the left (red) and right (blue) channel. The inset pictures show scheme of a DC and an example of waveguides writing order for $N$ = 5 scans. At the top of the scans, the numbers from 0 to 9 show the writing order, which are also connected by a line sequentially. (b) Same statistics for MZIs. In this case reflectance is affected by the internal phase $\delta \phi$, which is determined by the homogeneity of the waveguides. The inset picture shows scheme of an MZI.}
\label{fig:fig5}
\end{figure}

\begin{figure*}[ht!]
\centering
\includegraphics[width=1.0\linewidth]{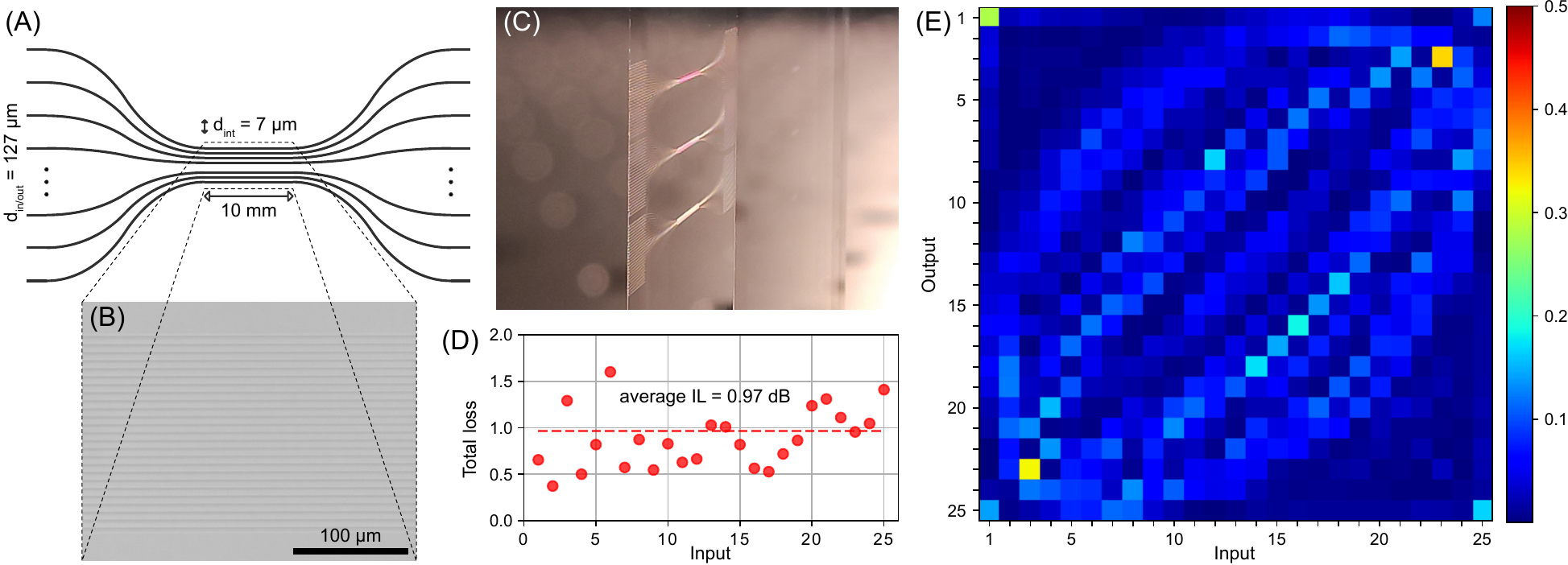}
\caption{(a) Scheme of the 25-channel interferometer. (b) Microphotograph of the interferometer's straight section (interaction region). (c) Photograph of a sample containing three interferometers. (d) IL of the interferometer depending on the input channel. The dash line indicates the average IL of 0.97 dB. (e) Experimentally obtained $25\times25$ transfer matrix of the interferometer.}
\label{fig:fig6}
\end{figure*}

The multiscan waveguides take quite a long time to write. For example a 5~cm long waveguide consisting of 19 consecutive scan cycles requires $\approx$ 15 minutes of fabrication time. When waveguides of an optical circuit are written sequentially, any changes in the writing conditions over time (instability of the laser parameters, environmental conditions, etc.) negatively affect the uniformity of the circuit. The other approach to the problem of writing a large circuit takes a different path: applying the multiscan-with-a-shift method to the entire circuit, rather then to the separate waveguides. We call this approach a parallel waveguide writing procedure. We performed a comparative study of sequential and parallel writing procedures targeting the repeatability of the optical parameters in arrays of identical photonic structures. We fabricated sets of 10 DCs and 10 MZIs with the same parameters, where the waveguides in each element were written sequentially and in parallel. The DC repeatability statistics are shown in Fig.~\ref{fig:fig5}a, where the inset image illustrates an example of the scanning order for $N = 5$. The average value of DC reflectance  is 35.7 $\pm$ 1.9\% for sequential writing procedure and 34.1 $\pm$ 0.9\% for parallel writing, which indicates an almost two-fold improvement in repeatability if parallel writing is employed. The same statistics for the MZI samples are shown in Fig.~\ref{fig:fig5}b, where the reflectance values are 45.4 $\pm$ 4.4\% on average for sequential writing and 7.4 $\pm$ 2.6\% for parallel writing. The MZI reflectance is influenced by the internal phase difference $\delta \phi$ in the interferometer arms, which is determined by the homogeneity of the waveguides. For 7 mm long arms this difference is on average 0.4$\pi$ for sequential writing and 0.1$\pi$ for parallel writing, which indicates a significant uniformity improvement.

\subsection{\label{sec:E6}Low-loss 25-channel interferometer}

As an exemplary application demonstrating the capabilities of the multiscan writing technology developed, we fabricated and characterized a 25-channel static interferometer in a 5~cm long sample. The interferometer scheme is shown in Fig.~\ref{fig:fig6}a. The interferometer is implemented as an array of coupled waveguides with a 7~$\mu$m pitch between the waveguides in the interaction region, which is 10~mm long (see the microphotograph in Fig.~\ref{fig:fig6}b). The fan-in/out sections include waveguides with a minimal radius of curvature equal to 60~mm. The input and output channel spacing is 127~$\mu$m matching the standard fiber arrays. A photograph of the sample with three 25-channel interferometers is shown in Fig.~\ref{fig:fig6}c. We injected laser radiation into each input channel using a single fiber, and measured the power from each output channel using a fiber array and digital power meters. The dependence of the total IL on the input channel is shown in Fig.~\ref{fig:fig6}d and the average value equals to 0.97~dB. The measured $25\times25$ transfer matrix of the interferometer is shown in Fig.~\ref{fig:fig6}e.

\section{\label{sec:level1}Discussion and Conclusion}

The chip IL is one of the primary limiting factors for demanding applications like scaling up linear-optical quantum computing. Interestingly, if the PL is low enough, the CL with optical fibers dominates in the total IL value. Optimization of the waveguide shape and writing parameters allowed us to increase the overlap above 98.8\%, which in turn resulted in fairly low CL of 0.2 dB per facet. The PL were as low as 0.07 dB/cm, which is on par with the previously reported record value \cite{Tan22}. 

We found the decrease in the insertion loss with larger writing depth interesting. It is possible that due to the narrow pulse energy band the type-I waveguide formation regime at low depths (see Fig.~\ref{fig:fig1}) exhibits transition to the type-II regime during multiscan process. The positive aspects of the influence of spherical aberrations on properties of waveguides were also previously known \cite{FERREIRA2021, Ross-Adams2024}. Spherical aberrations elongate the type-I waveguide height to $\sim 10$~$\mu$m at depths of 700 $\mu$m and deeper. These can be used to create similar waveguides for the telecom wavelengths \cite{Cai2022review}. For short wavelengths, on the contrary, it is necessary to make the heights of the waveguides smaller $\sim 3-4$~$\mu$m, for which focusing lenses with a larger NA $>$ 0.55 can be used.

Despite the fact that we properly simulated the writing regimes discussed above, the results still leave room for further writing regime optimization. For an improvement of CL, the close to circular shape of multiscan waveguide can be considered \cite{Sun2022, Wang2024}. Other shapes can be beneficial in term of minimizing BL \cite{Wang2024}. However, widening of the waveguides may lead to spontaneous formation of microcracks during sample polishing \cite{Lee2021}. Thermal annealing \cite{Witcher13, Arriola13, Amorim19} or half-scan technique \cite{Sun2020} could be helpful for relieving induced cumulative stress. An equally important result of the work is the improvement in the homogeneity of waveguides during parallel writing. This can lead to increased accuracy of manufactured circuits and devices.

An obvious disadvantage of multiscan waveguides is the low effective writing speed. However, as the writing speed study has shown, the fabrication process can be accelerated 4-6 times with moderate sacrifices in the CL. It is worth mentioning that in certain cases fewer scans may be sufficient \cite{Ehrhardt23}. 

Finally, we achieved the total IL of the 25-channel interferometer below 1 dB, which is the lowest value among previously reported \cite{Albiero2022, Pont2022, Pont2022_2, Albiero2024, Tan22}. We hope that the proposed multiscan waveguide writing approach will enable the fabrication of low-loss photonic integrated circuits with great potential for quantum applications.

\begin{acknowledgments}
The work was supported by Russian Science Foundation grant 22-12-00353 (https://rscf.ru/en/project/22-12-00353/) in part of the development of the multiscan waveguides. The work was supported by Rosatom in the framework of the Roadmap for Quantum computing (Contract No. 868-1.3-15/15-2021 dated October 5, 2021 and Contract No.P2154 dated November 24, 2021) in part of fabrication and characterization of 25-channel interferometer. S.P.K. and A.A.K acknowledge support by the Ministry of Science and Higher Education of the Russian Federation on the basis of the FSAEIHE SUSU (NRU) (Agreement No. 075-15-2022-1116). S.A.Z. acknowledges support by the Foundation for the Advancement of Theoretical Physics and Mathematics ``BASIS'' (22-2-2-26-1).
\end{acknowledgments}

\appendix

\section{\label{app:setups_schemes}Schemes of the FLW and characterization setups}

\begin{figure}[ht!]
\centering
\includegraphics[width=1.0\linewidth]{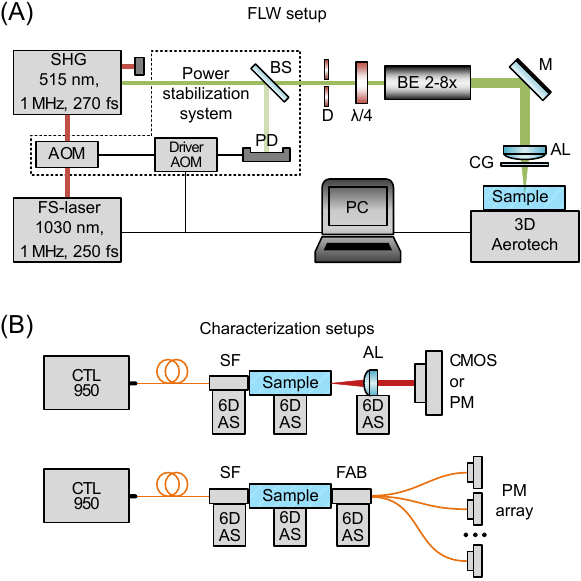}
\caption{(a) The FLW setup scheme. SHG -- second harmonic generator, AOM -- acousto-optic modulator, $\lambda /4$ --quarter-waveplate, BS -- beamsplitter, PD -- photodiode, BE -- beam expander, M -- mirror, AL -- aspheric lens ($NA = 0.55$), CG -- cover glass, 3D Aerotech -- automated three-axis positioning system. (b) Waveguide transmittance and mode characterization setup scheme (top). The 25-channel interferometer characterization setup scheme (bottom). CTL950 -- diode laser, SF -- single fiber, 6D AS -- 6-axis alignment stage, AL -- aspheric lens ($F$ = 7.5 mm), FAB -- fiber array block, PM -- power meter, CMOS -- beam profiler.}
\label{fig:figA1}
\end{figure}

The FLW setup scheme is shown in Fig.~\ref{fig:figA1}a. 

The waveguides transmittance and mode characterization setup scheme is shown in Fig.~\ref{fig:figA1}b (top). Transmittance was determined as the ratio between a measured output power (after collimation) and input power (before the sample) as $T = P_{out}/P_{in}$. IL was calculated as $IL = -10 \times lg(T)$, which can be decomposed into the sum of loss values with different origins $IL = CL + PL \times L + BL \times L_b$. The CL includes the Fresnel reflection loss. BL is an additional bending loss for curved waveguides with a length $L_b$. 

This setup was also employed for the characterization of DCs and MZIs. The reflectance of the DCs and MZIs was calculated as the ratio between the measured output power from the R channel and the sum of the powers from both the R and T channels as $R = P_{out R}/(P_{out R} + P_{out T})$. The internal phase difference in the MZI arms, designated as $\delta \phi$, was evaluated from its transfer matrix, taking into account the measured average reflectance of the DCs:

\begin{equation*}
MZI = DC(R) \times Phase(\delta \phi) \times DC(R),
\end{equation*}

\begin{equation*}
DC(R) = 
\begin{pmatrix}
\sqrt{R} & i \sqrt{1 - R} \\
i \sqrt{1 - R} & \sqrt{R}
\end{pmatrix},
\end{equation*}

\begin{equation*}
Phase(\delta \phi) = 
\begin{pmatrix}
e^{i \delta \phi} & 0 \\
0 & 1
\end{pmatrix}.
\end{equation*}

The 25-channel interferometer characterization setup scheme is shown in Fig.~\ref{fig:figA1}b (bottom). IL of the 25-channel interferometer was determined as the ratio between sum of the measured output powers and input power (before the sample) as $IL = -10 \times lg(\sum P_{out}/P_{in})$, which can be decomposed into the sum of loss values $IL = 2 \times CL + PL \times L + BL \times L_b$.

\section{\label{app:simulation}Simulations of the fiber coupling efficiency with rectangular waveguides}

\begin{figure*}[h!]
\centering
\includegraphics[width=0.9\linewidth]{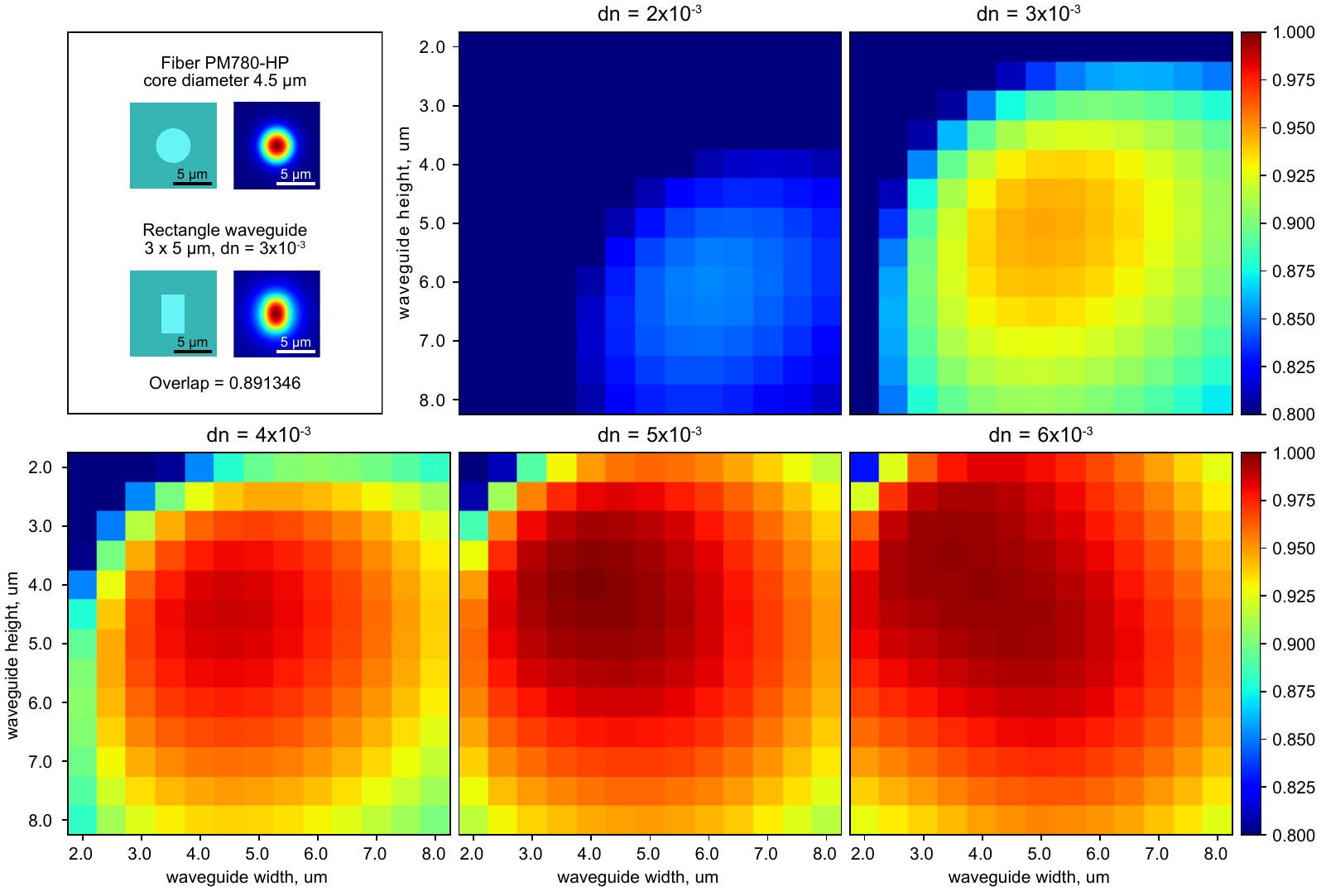}
\caption{Example of a simulation task: waveguide and fiber cores, their modes and the calculated
overlap. Simulation results: mode overlaps plot for rectangular waveguides of different sizes from 2~$\mu$m to 8~$\mu$m depending on the refractive index contrast in range $2 - 6 \times 10^{-3}$.}
\label{fig:figA2}
\end{figure*}

\begin{figure*}[h!]
\centering
\includegraphics[width=0.7\linewidth]{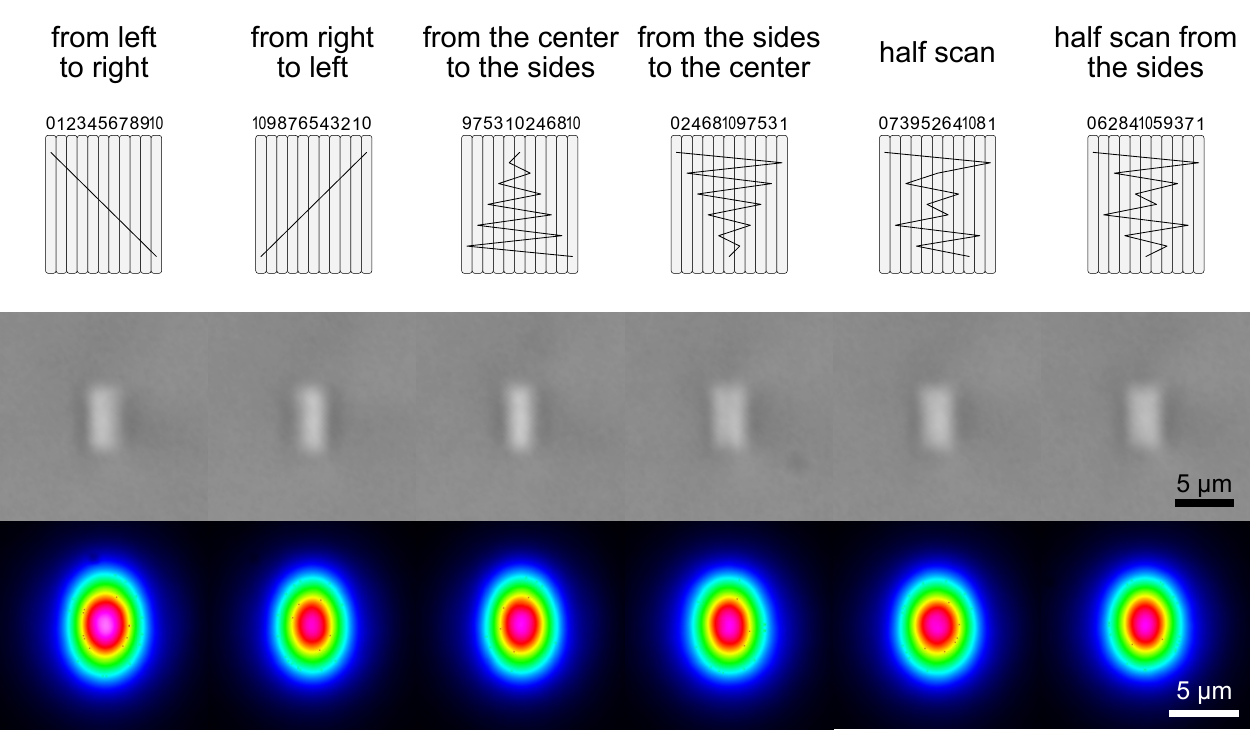}
\caption{Comparison of writing order in multiscan waveguides with $N$ = 11: schemes of ordering principles under consideration, facet view micrographs of waveguides in DIC regime and mode profiles of waveguides for H-polarization.}
\label{fig:figA3}
\end{figure*}

The considered PM780-HP fiber core diameter was taken from the manufacturer (Coherent) open source specifications and was equal to 4.5~$\mu$m. The fiber core refractive index was selected $n_{fiber}$ = 1.45666 such that its mode size is equal to the experimentally measured $MFD_{fiber}$ = 7.8~$\mu$m. The fiber and waveguides cladding refractive index was selected $n_{clad}$ = 1.45150 to be the same as pure fused silica glass for 920 nm. All core widths were increased by a minor amount of 10 nm to avoid encountering mode superposition due to degeneracy in perfectly circle and square cores. This had no effect on the resulting value of the overlap. Rectangular waveguides heights and widths from 2~$\mu$m to 8~$\mu$m and $\Delta n$ in range from $1 \times 10^{-3}$ to $6 \times 10^{-3}$ were considered. Simulation results are shown in Fig.~\ref{fig:figA2}.

\section{\label{app:writing_order}Writing scans order in multiscan waveguides}

In multiscan waveguides, the writing order of individual consecutive writing scans affects the properties of the resulting waveguide. To choose the most suitable one, we considered the following principles: "from left to right", "from right to left", "from the center to the sides", "from the sides to the center", the writing order in which is clear from their names. Also we considered the "half scan" \cite{Sun2020} and its slightly modified version "half scan from the sides" principles. All scans were written in the same direction. The schemes of the scans order principles for $N$ = 11 are shown in Fig.~\ref{fig:figA3}. At the top of the scans, the numbers from 0 to 10 show the writing order, which are also connected by a line sequentially. We have written a waveguide for each principle and compared their characteristics: homogeneity, refractive index contrast, mode profile, transmittance. Facet view micrographs of waveguides in DIC regime and mode profiles of waveguides for H-polarization are shown in Fig.~\ref{fig:figA3}. According to the results, the principle "from the center to the sides" was chosen as the most suitable and was used for experiments in the main text.



\bibliography{apssamp}

\end{document}